\documentclass{aa}
\usepackage{graphicx,natbib}
\bibpunct{(}{)}{;}{a}{}{,}

\def\mnras{MNRAS}

\begin{document}

\sloppypar

\title{Aperiodic optical variability of intermediate polars -- cataclysmic variables with truncated accretion disks}

\author{Revnivtsev M. \inst{1,2}, Burenin R.\inst{2}, Bikmaev I. \inst{3},
  Kniazev A. \inst{4,5}, Buckley D.A.H. \inst{4,5}, Pretorius M.L.,
  \inst{6,4,5}, Khamitov I. \inst{7}, Ak T. \inst{7,8}, Eker Z.\inst{7},
  Melnikov S.  \inst{3}, Crawford S. \inst{4}, Pavlinsky M. \inst{2}}


\institute{
  Excellence Cluster Universe, Technische Universit\"at M\"unchen,
  Boltzmannstr.2, 85748 Garching, Germany
  \and
  Space Research Institute, Russian Academy of Sciences,
  Profsoyuznaya 84/32, 117997 Moscow, Russia
  \and
  Kazan State University, ul. Kremlevskaya 18, Kazan, Russia
  \and
  South African Astronomical Observatory, PO Box 9, 7935 Observatory, Cape
  Town, South Africa.
  \and
  Southern African Large Telescope Foundation, PO Box 9, 7935 Observatory,
  Cape Town, South Africa
  \and
  European Southern Observatory, Alonso de Cordova 3107, Santiago, Chile
  \and
  TUBITAK National Observatory, 07058, Antalya, Turkey
  \and
  Istanbul University Faculty of Science Department of Astronomy and
  Space Sciences, 34119 University, Istanbul, Turkey
}
\date{}
\authorrunning{Revnivtsev et al.}
\titlerunning{Aperiodic optical variability of intermediate polars}

\abstract{We study the power spectra of the variability of seven intermediate
  polars containing magnetized asynchronous accreting white dwarfs, XSS J00564+4548,
  IGR J00234+6141, DO Dra, V1223 Sgr, IGR J15094-6649, IGR J16500-3307 and
  IGR J17195-4100, in the optical band and demonstrate that their
  variability can be well described by a model based on fluctuations
  propagating in a truncated accretion disk. The power spectra have breaks
  at Fourier frequencies, which we associate with the Keplerian frequency of
  the disk at the boundary of the white dwarfs' magnetospheres. We propose
  that the properties of the optical power spectra can be used to deduce the
  geometry of the inner parts of the accretion disk, in particular: 1)
  truncation radii of the magnetically disrupted accretion disks in
  intermediate polars, 2) the truncation radii of the accretion disk in
  quiescent states of dwarf novae. }

\maketitle

%

\section{Introduction}

Aperiodic variations in the intensity of many astrophysical sources carry
a lot of information about physical processes in and around them, which is
complementary to and completely independent of all other types of astrophysical
information. It was noticed already long ago that accreting sources
typically demonstrate flux variability on a wide range of time scales, from
milliseconds and seconds for galactic compact objects (e.g.
\citealt{linnel50} for accreting white dwarfs, \citealt{rappaport71,oda74}
for X-ray binaries) to months and years for accreting supermassive black
holes in active galactic nuclei \cite[e.g.][]{zaitseva69}.  Aperiodic flux
variations of X-ray emitting compact sources were discovered since the 1970s
and have been extensively studied since then. Typically these aperiodic flux
variations have a power spectrum (square of Fourier transform of the light
curve) that can be described as a power law with a slope $P\propto f^{-1
  ... -1.5}$, sometimes with flattening at low Fourier frequencies and
steepening at high Fourier frequencies. It was recognized that photons,
which contain this flux variability, originate in the innermost regions of
the accretion flow. While for X-ray accreting sources it was quite clear
from the beginning (simply from energy contained in the variable part of the
X-ray radiation), for accreting white dwarfs it was shown with the help of
eclipse mapping \cite[e.g.][]{bruch92,bruch96,baptista04}.

However, until recently the origin of the these variations remained unclear.
This variability possesses a set of properties that was not easy to explain
in early models. For example, the shot-noise model (see e.g.
\citealt{terrell72}) proposed that the variability originates as a result of
additive summation of randomly occurring exponential ''shots''. This model
explained the shape of power spectra of some X-ray binaries
\cite[e.g.][]{terrell72,kraicheva99}, but failed to reproduce very essential
observational properties: 1) the extremely wide range of variability time
scales, on which the variability has a self-similar power spectrum in some
cases \citep{churazov01}, 2) the linear dependence of the amplitude of flux
variability on the time average flux of the source \citep{uttley01}, 3)
the log-normal distribution of instantaneous values of the flux and 4) the behavior
of the frequency dependent phase lags between lightcurves in different energy
bands \citep{miyamoto88,kotov01}.

Now it is widely accepted that the most appropriate model of the flux
variability of accreting sources is the model of propagating fluctuations
\citep{lyubarskii97,churazov01}.  In this model the variability of observed
flux originates as a transformation of variability of the instantaneous mass
accretion rate in the innermost regions of the accretion flow (where the
majority of the emission is created) into a variability of the emergent luminosity. In turn, the variability
of the mass accretion rate in the innermost region is a result of
propagation (in the flow going inward) of variability created at all outer
radii of the extended accretion disk. All inner regions of the disk create
additional modulations to the instantaneous mass accretion rate on top of
already existing longer time scale variability, transported to this radius
from outer parts of the disk. The combined action of all radii of the disk
creates power spectra with relatively shallow slopes ($-1$ \dots $-1.5$)
 (see e.g.
\citealt{lyubarskii97,churazov01}).  This relatively simple construction
allows us to explain all the major observational characteristics of flux
variability (see more details in
\citealt{churazov01,kotov01,uttley01,revnivtsev08}).

In the framework of this model one can make very specific predictions for
the shape of the power spectra of flux variability of accretion flows.  For
example, the accretion disk with an inner boundary should have a break in
the power density spectrum of its time variability, approximately at the
Fourier frequency, which corresponds to the typical time scales of
variability introduced into the flow at the inner radius of the disk.

Accretion-powered X-ray pulsars and asynchronous magnetic white dwarfs
(intermediate polars) have magnetic fields strong enough to disrupt the
inner parts of the accretion disks.  Thus the the fastest variability
timescales associated with the innermost regions of the disk should be
absent or reduced in their power spectra.

This assumption was recently explored by \cite{revnivtsev09} in the
X-ray energy domain, who analyzed a large sample of magnetized accretion-
powered objects (X-ray pulsars and an intermediate polar) and showed that
the variabilities of all these sources indeed have a break in their power spectra
at the anticipated Fourier frequencies. The frequency of these breaks
corresponds to the the inner disk Keplerian frequency at the boundary of the
central object's magnetosphere.  The conclusion was strongly supported because
the frequency of the break varied during the giant outbursts
of accretion-powered X-ray pulsars, in accordance with the simple prediction
of how the radius of the dipole magnetosphere of the central object (and
consequently the disk Keplerian frequency at this boundary) should vary with
the mass accretion rate (X-ray luminosity).  Among other things, this study
opens a new possibility to measure the properties of magnetospheres of
compact objects via the study of their flux variability. In a lot of cases this
approach is much easier than trying to measure the cyclotron absorption
lines in spectra of pulsars or polarization of photons originated in the
magnetized accretion curtains in intermediate polars.

We present results of our pilot study of aperiodic
variability of some asynchronous magnetized accreting white dwarfs
(intermediate polars) with accretion disks ({\sl this is an essential
  ingredient of binary systems in our sample, because sometimes
  asynchronous magnetized accreting white dwarfs might be stream-fed rather
  then disk-fed} , see e.g. \citealt{hellier02}) in the optical
bands (B,V,g',R) and show the applicability of the method.

Observational data for our study was collected with the help of the
Russian-Turkish telescope RTT150, the Southern African Large Telescope
(SALT), 0.76m and 1m telescopes of South African Astronomical Observatory.

\section{Observational data}
\label{sec:data}

Power spectra of variability of optical light curves of accreting intermediate polars
were obtained via Lomb-Scargle periodograms \citep{lomb76,scargle82}, averaged over all available uninterrupted datasets. As the typical length of datasets is approximately several hours, the
lowest Fourier frequency probed by our power spectra were $\sim10^{-4}$ Hz, while the highest
frequencies depend on the time resolution used. In Table 1 we present the list of sources, used in 
our analysis. Optical magnitudes of binaries are approximate, because of significant long term flickering, present in all systems. Photometry of SALT is subject to additional uncertainty because of the method of observations (see below). 

\begin{table*}[htb]
\caption{Optical observations of accreting white dwarfs used for
  construction of power spectra.}

\renewcommand{\arraystretch}{1.4}
\begin{tabular}{l|c|c|c|c|c|c}

Target & Instr.& Dates&Mag&Time res., sec& Exp., ksec & Spin Period, s\\
\hline
XSS J00564+4548&RTT150&2005.10.19-23,26,28&14.9(V)&17-128&71.6&465\\
               &      &2005.12.05,06        &14.6(R)&      &    &   \\
IGR J00234+6141&RTT150&2005.09.10-11,12.01&16.7(R)&60-70&37.6&561\\
               &      &2005.12.10-11,2007.09.03&16.7(R)&   &   \\
DO Dra    & RTT150&2007.04.24,27,28 &14.7(R)& 1 &48& 529\\
          &       &2008.05.24-26,06.03&14.8($g'$)&1-8&39&\\
V1223 Sgr & SALT  &2008.05.13,2008.06.12   &$\sim$13.3(B)&0.1 & 2.9&745\\
IGR J15094-6649&SAAO, 1m&2008.04.26-29&14.6(V)&7-8&14.9&809\\
IGR J16500-3307&SAAO, 1m&2008.04.27-28,05.04&15.9(V)&8-10&18.4&598$^*$\\
               &SAAO, 0.76m&2008.08.03-05      &15.9(V)&  &\\
IGR J17195-4100&SAAO, 0.76m&2008.08.5-10&15.2(V)&8&32.3&1139$^*$\\
\end{tabular}
\bigskip
\begin{list}{}
\item $^*$ --- here we use the optical pulsation period, which also may be
  the beat frequency instead of the true spin frequency of the white dwarf
\end{list}
\end{table*}

\subsection{RTT150}

Optical observations of the intermediate polars XSS~J00564+4548,
IGR~J00234+6141 and DO~Dra (=YY Dra) were carried out with 1.5-m
Russian-Turkish Telescope (RTT150) at T\"{U}B\.ITAK National Observatory
(TUG), Bak\.irl\.itepe mountain, Turkey. The objects were observed as a part
of a long-term program of study of aperiodic optical variability of cataclysmic variables (CVs).

We used a low-readout noise back-illuminated 2$\times$2\,K Andor DW436 CCD in
combination with $B$, $g'$, $V$, or $R$ filters, mounted in F/7.7 Cassegrain
focus of the telescope. The measurements with $>1$~s time resolution were
obtained from the series of direct images taken with only a small CCD
subframe near the object. The bias was subtracted and flat-field correction
was applied for all images. The subframe included also at least one bright
non-variable reference star, used for differential photometry. The use of
differential photometry allowed us to diminish the influence of atmospheric
turbulence on the resulting measurements, which is extremely important for
aperiodic variability studies. Data reduction and photometric measurements
were made with IRAF\footnote{http://tucana.tuc.noao.edu/} and our own
software.

The measurements with $\approx1$~s time resolution were made with the
setup where the narrow strip of the CCD was binned in its width into only
one pixel, so that only one line of the data was read out of the CCD in each
exposure. That was done to diminish the CCD read-out time. This
setup also allowed us to make the differential photometric measurements, since
the bright non-variable bright reference star was also observed in another
part of this strip. We checked that these differential photometric
measurements effectively eliminated the influence of atmospheric turbulence
even at these higher frequencies. The detailed description of this setup and
appropriate calibration measurements can be found in Burenin et al. (in prep.).

\subsection{SAAO}

Light curves of three intermediate polars IGR J15094-6649, IGR J16500-3307
and IGR J17195-4100 were obtained with the SAAO 1-m and 0.76-m telescopes
and the UCT CCD (this is a frame-transfer CCD, implying that there is no
dead time between exposures). Detailed description of these data is
presented in \cite{pretorius09}.

\subsection{SALT}

High time-resolution observations of V1223\,Sgr were taken at the prime
focus of the newly available 10-m class {\em Southern African Large
  Telescope} \citep[SALT;][]{Buck06,Dono06} during its performance
verification phase on 2008 May 13 and June 12.  Observations were done with the
imaging camera \citep[SALTICAM;][]{Dono03,Dono04} in `slot mode'
\citep{Dono06}.  The SALTICAM is a CCD mosaic of two 2048$\times$4102 pixels
CCDs with full image size of about 9.6$\times$9.6 arcmin, but the nominal
science field is 8 arcmin.  In 'slot mode', a mask with a slot of the width of 20
arcsec (144 unbinned rows) is projecting on the CCDs. Only the central 72 rows
receive all the light.  The slot extends across the full 8 arcmin in the horizontal
direction. After each exposure, only 144 rows are transferred across the
frame-transfer boundary, which shortens the vertical clocking overhead to 14 ms.
The instrument thus provided a sequence of images 8 arcmin long, but only 20
arcsec wide.  These images were split into four distinct sections,
corresponding to the four readout amplifiers of the two CCDs.

Total tracking time was $\sim$2700 sec on May 13 and $\sim$1300 sec on
June 12.  The observations were made with the $B$ filter. In order to
facilitate the shortest possible exposures, 6$\times$6 binning, or 1.2
arcsec/pixel was used.  The exposure time for each image was 112 ms (the time to
move each image behind the mask, during which image smearing takes place, is
14 ms).  The observed data were all referenced to the nearby star of similar
magnitude, 63 arcsec to the east and 7 arcsec south of the target star.  The
comparison star and the target star were imaged within the same amplifier.
No flat-field calibration frames were obtained (the calibration system was
not available at the time of the observations) and the frames were simply
bias-subtracted with overscanned pixels from each row.

The data reduction of ``slot mode'' was done with the the {\tt PySALT} user
package.  This tool is primarily written in python/PyRAF with some
additional IRAF code\footnote{ See
  http://www.salt.ac.za/science-support/salt-data-reduction/pysalt-users-package/
  for more information.}.

Analysis of the lightcurves obtained with SALTICAM in the described mode
revealed problems in power spectra of celestial sources at Fourier
frequencies higher than $\sim$1Hz, while at lower frequencies the results
remained reliable. Therefore in our subsequent analysis we have not
considered power spectra of the considered source at Fourier frequency
higher than 0.7 Hz.

Power spectra of V1223 Sgr obtained in different observations were
virtually indistinguishable from each other, therefore we averaged
them.

\section{Discussion}

\subsection{Power spectra of optical emission of CVs}
\label{sec:pow}
Optical emission of white dwarfs\footnote{We emphasize
  that we consider here only magnetic cataclysmic variables {\sl with
    accretion disks} which might not always be the case even for
  asynchronous magnetic CVs, taking into account the existence of stream fed
  systems like V2400~Oph, see, e.g. \citealt{hellier02}.} consists of two parts: one is provided by the internal energy dissipation inside the disk and another is a reprocessed EUV or X-ray emission of white dwarf polar caps.

In the framework of the model of
propagating fluctuations we predict that the variability of the optical
light in intermediate polars will be similar to that of the X-rays. There
are two main scenarios:
\begin{enumerate}
\item If the optical emission of the system originates from the reprocessed
  X-rays, the X-ray and optical variability should be similar by construction.
\item If the optical light is due to internal heating of the extended accretion disk, we
  should still expect to observe similar power spectra of optical and X-ray
  light curves. That is because in both cases the flux variability is caused
  by the same fluctuations of the mass accretion rate. These fluctuations first
  affect the optical emission of the disk and after that, when this same
  matter falls down onto the polar caps of the white dwarf, it changes the
  X-ray emission in the accretion column.
\end{enumerate}
In both cases power spectra of X-ray and optical variability of the CVs
should be similar, and we can study the shape of the power spectrum of the X-ray
light curve via observations of the optical variability of these CVs.

The majority of white dwarfs are faint in X-rays (typical fluxes are less
than on the order of few photons/1000 sq.cm./sec), therefore it is very
difficult to observe them in X-rays. However, they can be more easily
observed in the optical band. Therefore the study of optical variability in
accreting white dwarfs can provide important diagnostics of the geometry of
the inner parts of the accretion flow. In particular, we can immediately
apply the diagnostics

\begin{itemize}
\item to make estimates of the magnetic moment of white dwarfs.  In the
  simplest model of dipolar magnetosphere of the white dwarf, the dependence
  of the inner disk radius on the mass accretion rate in the disk is
  determined mainly by the white dwarf magnetic moment. Therefore
  measurements of power spectra of white dwarf variability in different
  accretion states will allow us to make estimates of the magnetic field.
  The proposed approach might be more sensitive than the search for
  cyclotron emission lines in optical-NIR spectra of objects.
\item to make estimates of holes in the innermost regions of the
  accretion disk due to evaporation, predicted in some models of the
  quiescent accretion on white dwarfs \citep[e.g][]{meyer94}.
\end{itemize}

\begin{figure}
\includegraphics[width=\columnwidth]{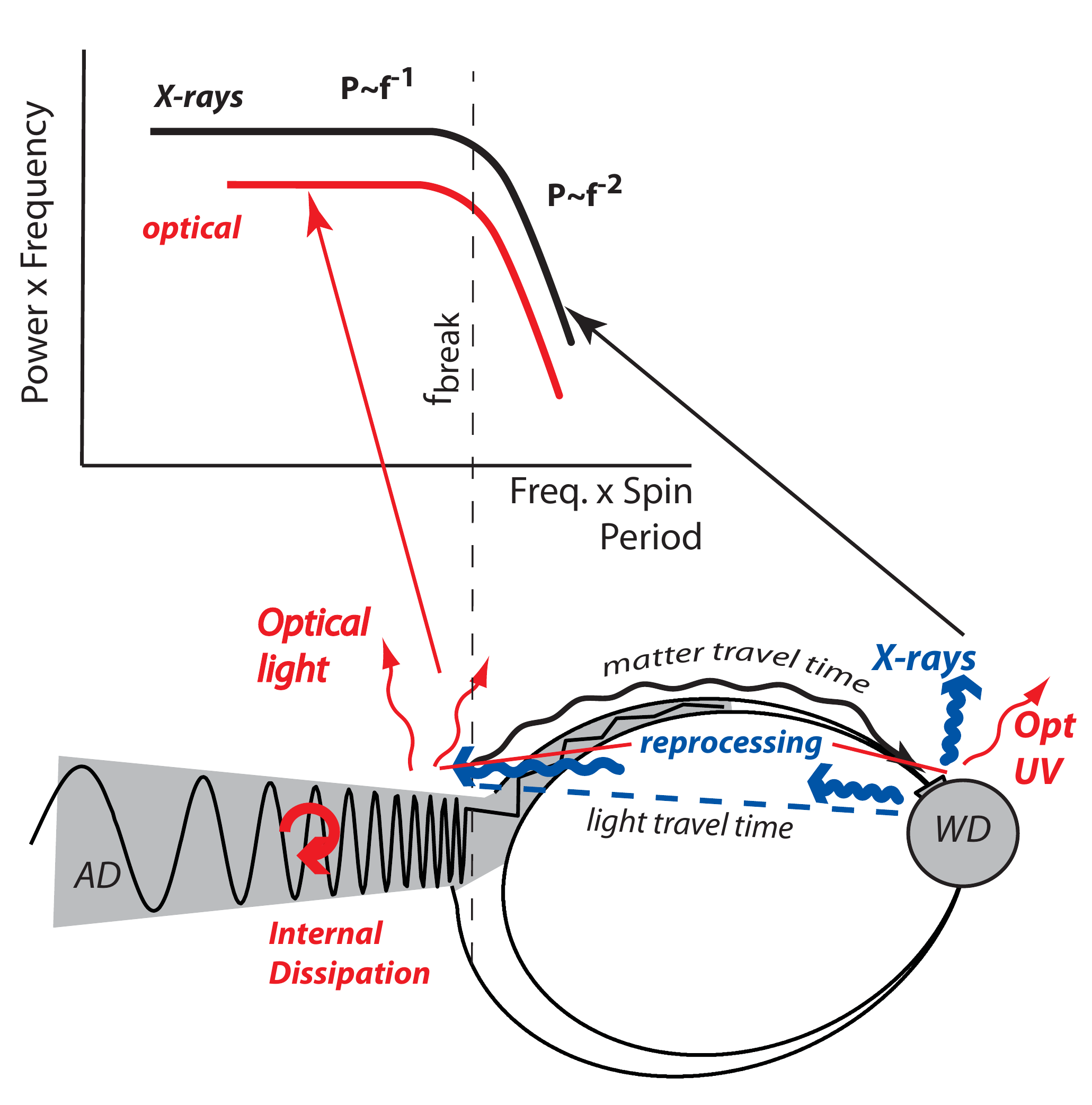}
\caption{Scheme of accretion flow in magnetized accreting white dwarfs}
\label{scheme}
\end{figure}

\subsection{Observed power spectra}

\begin{figure}
  \includegraphics[width=\columnwidth]{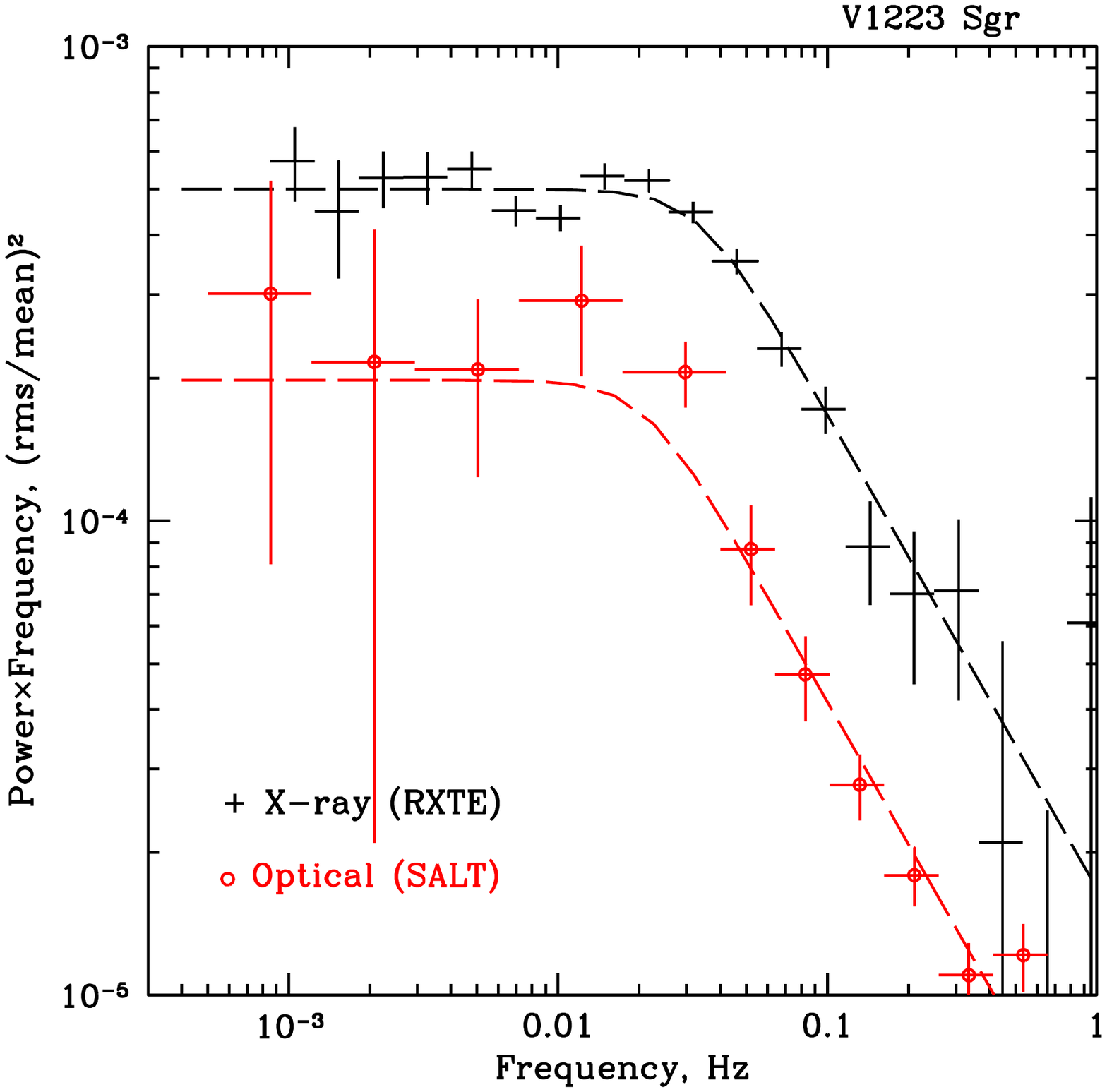}
  \caption{Power spectra of V1223 Sgr in X-ray and in optical spectral
    bands. Dashed curves denote the simple analytic model described in the text ( $P\propto f^{-1} (1+[f/f_0]^4)^{-1/4}$ ) fitted to the X-ray (black curve) and optical (red curve) power spectra.}
  \label{powers1223}
\end{figure}

The observations of one of the brightest intermediate polars, V1223 Sgr, agree
well with the considerations presented above. In Fig.~\ref{powers1223} we
present the power spectra of V1223 Sgr in optical and in X-rays
\citep[from][]{revnivtsev09}. The similarity of these power spectra is striking. We
fitted these power density spectra with a simple analytical function
$P(f)\propto f^{-1} (1+[f/f_0]^4)^{-1/4}$ (essential property whose is that
it has a slope $P\propto f^{-1}$ at frequencies $f<f_{0}$ and a slope $P\propto
f^{-2}$ at higher Fourier frequencies). The values of the break frequencies
in optical and in X-rays are: $f_{\rm break, opt}=(2.1\pm0.5)\times10^{-2}$
Hz, $f_{\rm break, X-ray}=(3.36\pm0.3)\times10^{-2}$ Hz (in order to determine the $f_{\rm break}$ values and their confidence intervals we used the $\chi^2$ minimization technique). The break
frequencies are compatible at the 2$\sigma$ level. There is marginal
evidence for a higher break-frequency in X-rays, but it may be
explained, for example, by a long-term difference in the accretion-rate of the source, 
since the optical and X-ray data were obtained in different epochs.

Assuming that the break frequency corresponds to the Keplerian frequency at
the boundary of the white dwarf magnetosphere \citep{revnivtsev09}, we can
estimate its radius as
$$
r_{\rm m}\sim1.6\times10^{9}\,\,{\rm cm}
$$
or $\sim2.8 R_{\rm WD}$ (here we adopted the mass of the white dwarf in this
system is $M_{\rm WD}=0.95 M_\odot$, \citealt{suleimanov05}). However, we
should keep in mind that due to a lack of exact understanding of the
formation of the break that this estimate is only accurate to within a
factor of a few.

\begin{table}
\caption{Parameters of the break frequencies of the power spectra of the studied sources.}
\begin{center}
\begin{tabular}{l|c}
Source & $f_{\rm break}\times$Period\\
\hline
XSS 00564+4548	&$0.7\pm0.2$\\
IGR J00234+6141	&$1.0\pm0.3$\\
DO Dra		&$0.5\pm0.1$\\
V1223 Sgr	&$16\pm4$\\
IGR J15094-6649	&$1.3\pm0.3$\\
IGR J16500-3307	&$1.5\pm0.2$\\
IGR J17195-4100	&$2.0\pm0.3$\\
\end{tabular}
\end{center}
\end{table}

\begin{figure}
\includegraphics[width=\columnwidth]{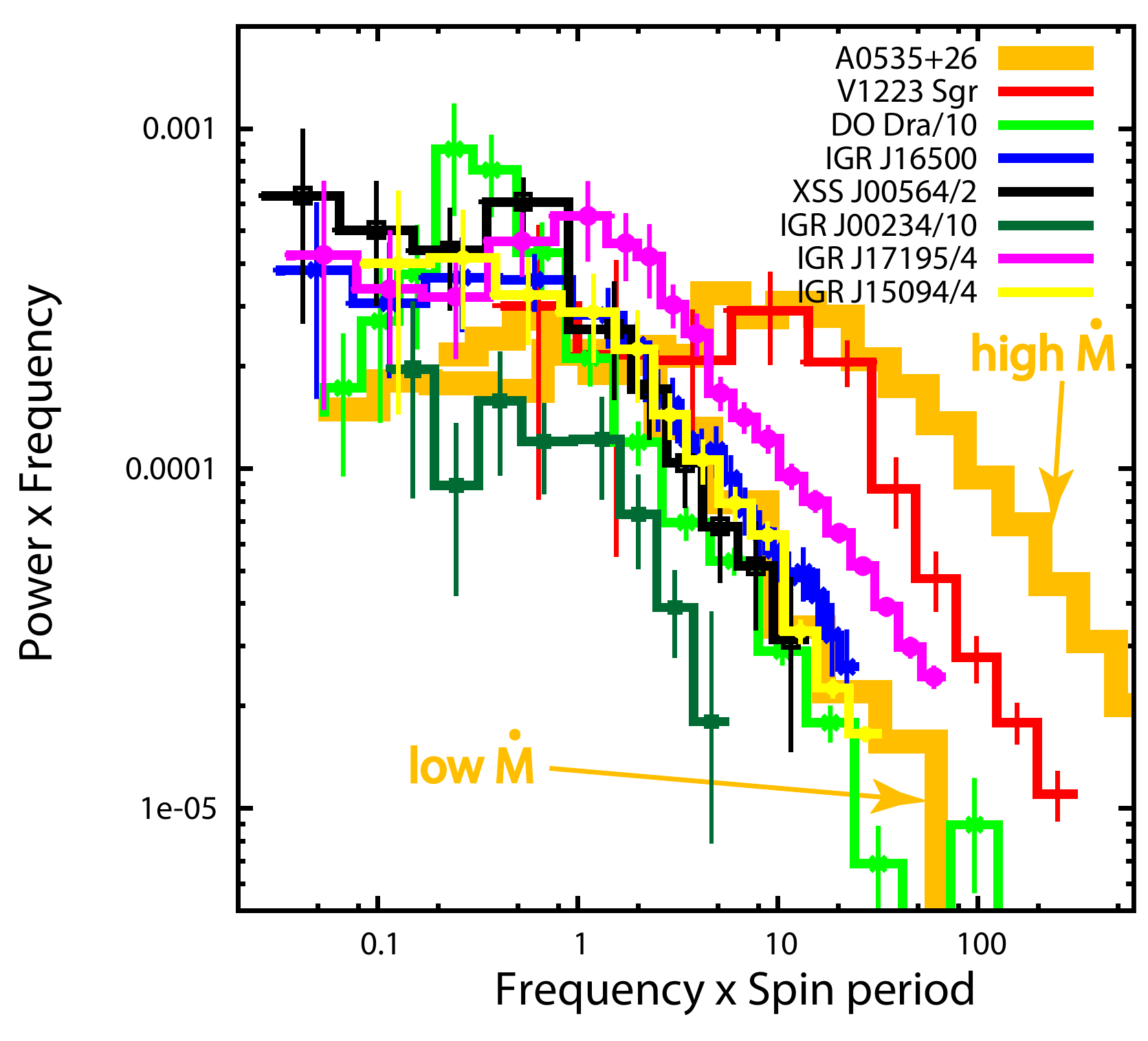}
\caption{Power spectra of optical variability of a sample of intermediate
  polars. Power spectra of DO Dra, XSS J00564+4548, IGR J00234+6141, IGR J17195-4100, 
and IGR J15094-6649 were scaled down for clarity by factors, written on the figure.   For comparison we present two power spectra of accreting X-ray
  pulsar A0535+26, collected during episodes of high and low mass accretion
  rates (thick solid histograms). See text for the details.}
\label{3powers}
\end{figure}

In Fig.~\ref{3powers} we present power spectra of optical variability of
seven objects --- XSS J00564+4548, IGR J00234+6141, V1223 Sgr, DO Dra, IGR
J15094-6649, IGR J16500-3307 and IGR J17195-4100, measured with data
described in the Sect.~\ref{sec:data}. 
The frequency axis of the power spectra was multiplied by the spin period of
the white dwarf magnetospheres. In the power spectra of a majority of these
objects we removed the peak (simply by removing measurements of a power at Fourier frequencies near the inverse of WD rotational period), which was caused by spin modulated variations of the optical
brightness of the sources. Fainter (and possibly wider) QPO peaks may be present
also in power spectra of objects without us detecting them. If such
QPO peaks reside near the break frequency, they could somehow contribute to
the observed sharpness of the break in some power spectra in
Fig.\ref{3powers}. However, the discussion of these possible QPO peaks is
beyond the scope of this paper.

In the framework of our model we predict
that on this plot all accreting magnetic white dwarfs, in which the
inner parts of the accretion disks are in corotation with white dwarf magnetospheres, will have
breaks in the power spectra at a value around unity, i.e. around the WD spin
frequency (see e.g. Fig.\ref{3powers} and Table 2). If the accretion disk continues 
significantly below the corotation radius, the break in the power spectrum should be observed at
a higher Fourier frequency.

For comparison, we also show in Fig.\ref{3powers} the power spectra of an
accreting X-ray pulsar, A0535+26, during its low and high luminosity state,
i.e.\ during the periods of its high and low mass accretion rates (from \citealt{revnivtsev09}). One can
see that the power spectrum of the X-ray emission of A0535+26 with a high mass
accretion rate continues to higher Fourier frequencies ($f\times {\rm Spin\,
  Period}>1$), while variability at lower frequencies ($f \times {\rm
  Spin\, Period}\la 1$) remained at almost the same level.

The difference at high frequencies can be explained as an addition of an
extra noise component, which is presumably generated in the annulus of
accretion disk between the radii corresponding to the magnetosphere size in
the high accretion rate (small radius) and low accretion rate (large
radius). This ring and therefore the associated variability were absent in
the state with low accretion rate.

We can see that the behavior of accreting WDs is qualitatively very similar
to that of accreting neutron stars. Power spectra of all IPs except for
V1223 Sgr indicate that their white dwarfs are close to corotation with the
inner edges of their accretion disks, 
while in V1223 Sgr the accretion disk
should be truncated at a radius smaller than that of corotation.  Simple
models predict a spin-up of the central object in this case. 

 This prediction
does not agree with the detection of spin down, reported for this system
approximately 20 years ago by \cite{vanamerongen87}, but this discrepancy
may be explainable. For example, the source may have started to spin up
since that time. Therefore additional observations are needed to investigate
this. Also, there may be a more complicated interaction of the accretion disk with
the white dwarf magnetosphere, in which the angular momentum of the white
dwarf is diminishing while the accretion disk extends significantly below
the corotation radius (corotation radius in this case is $\sim16R_{\rm
  WD}$). As an argument in favor of this scenario we can mention that there
are many of cases when actively accreting magnetic neutron stars do
have a spin-down, while in a simple picture of accretion the matter settling
to the surface of the star should always speed up the rotation (if the disk
is prograde).

\section{Summary and perspectives}

We demonstrated that high quality power spectra of optical and
X-ray variability of the intermediate polar, V1223 Sgr (accreting white
dwarf with a magnetosphere that disrupts the accretion disk at relatively
large distances from white dwarf surface), are very similar to each other.
Both power spectra have a very distinct break at a Fourier frequency which
we associate with the Keplerian frequency at the inner radii of the
truncated accretion disk. At higher Fourier frequencies the power spectra
steepen.

The similarity of the V1223 Sgr power spectra in the X-ray and the optical is expected
in both scenarios of the origin of optical emission --- if
optical emission is dominated by the reprocessing of X-rays and if optical
emission originates in the disk itself due to its internal heating. 
However, these two scenarios predict different lags between the
optical and X-ray emission of IPs. In the former case the optical emission
should lag the X-rays, in the latter it should lead the X-rays. This issue
can be studied using simultaneous optical and X-ray observations of the
source.

With a sample of magnetized white dwarfs we demonstrated that this
behavior may be universal for white dwarfs with truncated accretion disks,
like intermediate polars, with the break frequencies depending on the mass
accretion rate. This behavior is very similar to that of accreting neutron
stars with truncated accretion disks (e.g. accretion powered X-ray pulsars).

We suggest that the values of the break frequencies in power spectra of
accreting white dwarfs can be used to make estimates of the inner radii of
the truncated accretion disks. In turn, these can be used to make estimates of
the white dwarf magnetic fields.

We proposed a few tests of the outlined paradigm. In particular, for observations of
intermediate polars at different mass accretion rates our interpretation
predicts that the break frequency of their power spectra should increase
with an increasing mass accretion rate in the binary system.

\begin{acknowledgements}
We thank TUBITAK, IKI and KSU for partial supports in using the RTT150
(Russian-Turkish 1.5-m telescope in Antalya) with the project numbers 998,999.
  The authors thank Vadim Arefiev (IKI) and Marat Gilfanov (IKI, MPA) for
  providing us data of their observational sets; Denis Stetsenko, Roman
  Zhuchkov, Almaz Galeev (KSU), Alexey Tkachenko, Denis Denissenko, Nikolay
  Aleksandrovitch (IKI) for assistance in observations at RTT150. MR thanks
  Eugene Churazov and Koji Mukai for useful discussions. This research made use of data
  obtained from the High Energy Astrophysics Science Archive Research Center
  Online Service, provided by the NASA/Goddard Space Flight Center.  This
  work was supported by a grant of Russian Foundation of Basic Research
  07-02-01004, 07-02-01051, 07-02-00961, 08-08-13734,
  09-02-97013-p-povolzh'e-a, NSh-5579.2008.2, NSh-4224.2008.2 and program of
  Presidium of RAS ``The origin and evolution of stars and galaxies'' (P04).
  Some of the observations reported in this paper were obtained with the
  Southern African Large Telescope (SALT), a consortium consisting of the
  National Research Foundation of South Africa, Nicholas Copernicus
  Astronomical Center of the Polish Academy of Sciences, Hobby Eberly
  Telescope Founding Institutions, Rutgers University,
  Georg-August-Universitaet G\"ottingen, University of Wisconsin-Madison,
  Carnegie Mellon University, University of Canterbury, United Kingdom SALT
  Consortium, University of North Carolina -- Chapel Hill, Dartmouth
  College, American Museum of Natural History and the Inter-University
  Centre for Astronomy and Astrophysics, India.  We are grateful for the
  support of numerous people during the SALT PV phase.
\end{acknowledgements}

\end{document}